\documentclass[a4paper]{article}
\usepackage{graphicx}
\usepackage{onecolceurws}

\usepackage{booktabs} 
\usepackage{graphicx}
\usepackage{subcaption}
\usepackage{enumitem}
\usepackage[singlelinecheck=false]{caption}
\usepackage{tabularx}
\usepackage{comment} 
\usepackage{amsmath}
\numberwithin{figure}{section}
\usepackage{lipsum} 
\usepackage{hyperref}
\usepackage{color}
\usepackage{multirow}
\setlist{nosep}

\title{Early Fusion Strategy for Entity-Relationship Retrieval}
\author{Pedro Saleiro$^{1,2}$, Nata\v{s}a Mili\'{c}-Frayling$^{4}$, Eduarda Mendes Rodrigues$^{1}$, Carlos Soares$^{1,3}$}

\institution{$^1$FEUP, $^2$LIACC, $^3$INESC TEC, Universidade do Porto, Portugal\\$^4$School of Computer Science, University of Nottingham, United Kingdom\\{\texttt{\{pssc,eduarda,csoares\}@fe.up.pt, natasa.milic-frayling@nottingham.ac.uk }}}

\begin{document}
\maketitle

\begin{abstract}
We address the task of entity-relationship (E-R) retrieval, i.e, given a query characterizing types of two or more entities and relationships between them, retrieve the relevant tuples of related entities. Answering E-R queries requires gathering and joining evidence from multiple unstructured documents. In this work, we consider entity and relationships of any type, i.e, characterized by context terms instead of pre-defined types or relationships. We propose a novel IR-centric approach for E-R retrieval, that builds on the basic early fusion design pattern for object retrieval, to provide extensible entity-relationship representations, suitable for complex, multi-relationships queries. We performed experiments with Wikipedia articles as entity representations combined with relationships extracted from ClueWeb-09-B with FACC1 entity linking. We obtained promising results using 3 different query collections comprising 469 E-R queries.

\end{abstract}

%
%



\section{Introduction}
In recent years, we have seen increased interest in using online information sources to find concise and precise information about specific issues, events, and entities rather than retrieving and reading entire documents and web pages. Modern search engines are now presenting entity cards, summarizing entity properties and related entities, to answer entity-bearing queries directly in the search engine result page. Examples of such queries are ``Who founded Intel?" and ``Works by Charles Rennie Mackintosh". 

Existing strategies for entity search can be divided in IR-centric and Semantic-Web-based approaches. The former usually rely on statistical language models to match and rank co-occurring terms in the proximity of the target entity \cite{balog2012expertise}. The latter consists in creating a SPARQL query and using it over a structured knowledge base to retrieve relevant RDF triples \cite{heath2011linked}. Neither of these paradigms provide good support for entity-relationship (E-R) retrieval, i.e., searching for multiple unknown entities and relationships connecting them. Contrary to traditional entity queries, E-R queries expect tuples of connected entities as answers. For instance, ``Ethnic groups by country" can be answered by tuples $<$\textit{ethnic group}, \textit{country}$>$, while ``Companies founded by the creator of Star Wars" is expecting tuples of the format $<$\textit{company}, \textit{George Lucas}$>$. In essence, an E-R query can be decomposed into a set of sub-queries that specify types of entities and types of relationships between entities.

Recent work in E-R search followed a Semantic-Web-based approach by extending SPARQL and creating an extended knowledge graph \cite{yahya2016relationship}. However, it is not always convenient to rely on a structured knowledge graph with pre-defined and constraining entity types. For instance, search over transient information sources, such as social media \cite{liang2014time} or online news \cite{saleiro2016timemachine}, require more flexible approaches.  

We hypothesize that it should be possible to generalize the term dependence models to represent entity-relationships and achieve effective E-R retrieval without entity type restrictions. We propose a novel IR-centric approach using fusion-based design patterns for E-R retrieval from unstructured texts. We make the first step in that direction by presenting an early fusion strategy that consists in creating meta-documents for entities and entity-pairs (relationships) and then apply standard retrieval models.

In order to leverage information about entities and relationships in a corpus, it is necessary to create a representation of entity related information that is amenable to E-R search. In our approach we focus on sentence level information about entities although it can be applied to more complex methods for text segmentation. We use Wikipedia entity articles and entity-pairs occurrences from ClueWeb-09-B data set with FACC1 text annotations that refer to entities found in the text, including the variances of their surface forms. Each entity is designated by its unique ID and for each unique entity instance we created \textit{entity documents} comprising a collection of sentences that contain the entity. These context documents are indexed, comprising the \emph{entity index}. The same is done by creating \textit{entity-pair documents} and the \emph{entity-pair index}. These two indices enable us to execute E-R queries using an early fusion strategy with two different retrieval models, Language Models and BM25. The approach was tested on a reasonably large-scale scenario, involving 4.1 million unique entities and 71.7 M of entity pairs. 

\section{Related Work}

Li et al. \cite{li2012entity} were the first to study relationship queries for structured querying entities over Wikipedia text with multiple predicates. This work used a query language with typed variables, for both entities and entity pairs, that integrates text conditions. First it computes individual predicates and then aggregates multiple predicate scores into a result score. The proposed method to score predicates relies on redundant co-occurrence contexts.

Yahya et al. \cite{yahya2016relationship} defined relationship queries as SPARQL-like subject-predicate-object (SPO) queries joined by one or more relationships. They cast this problem into a structured query language (SPARQL) and extended it to support textual phrases for each of the SPO arguments. Therefore it allows to combine both structured SPARQL-like triples and text simultaneously. 

In the scope of relational databases, keyword-based graph search has been widely studied, including ranking \cite{yu2009keyword}. However, these approaches do not consider full documents as graph nodes and are limited to structured data. While searching over structured data is precise it can be limited in various respects. To increase the recall when no results are returned and enable prioritization of results when there are too many, Elbassuoni et al. \cite{elbassuoni2009language} propose a language-model for ranking results. Similarly, the models like EntityRank by Cheng et al. \cite{cheng2007entityrank} and Shallow Semantic Queries by Li et al. \cite{li2012entity}, relax the predicate definitions in the structured queries and, instead, implement proximity operators to bind the instances across entity types. Yahya et al. \cite{yahya2016relationship} propose algorithms for application of a set of relaxation rules that yield higher recall.

Web documents contain term information that can be used to apply pattern heuristics and statistical analysis often used to infer entities as investigated by \cite{conrad1994system}, \cite{petkova2007proximity} and \cite{rennie2005using}. In fact, early work by Conrad and Utt \cite{conrad1994system} proposes a method that retrieves entities located in the proximity of a given keyword. They show that a fixed-size window around proper-names can be effective for supporting search for people and finding relationship among entities. Similar considerations of the co-occurrence statistics have been used to identify salient terminology, i.e. keyword to include in the document index \cite{petkova2007proximity}.

Existing approaches to the problem of entity-relationship (E-R) search are limited by pre-defined sets of both entity and relationship types. In this work, we generalize the problem to allow the search for entities and relationships without any restriction to a given set and we propose an IR-centric approach to address it. 

\section{Entity-Relationship Queries}

E-R queries aim to obtain a ordered list of entity tuples $T_E=$ $<$$E_i, E_{i+1},..., E_n$$>$ as a result. Contrary to entity search queries where the expected result is a ranked list of single entities, results of E-R queries should contain two or more entities. For instance, the complex information need ``\textit{Silicon Valley companies founded by Harvard graduates}'' expects entity-pairs (2-tuples) $<$\textit{company}, \textit{founder}$>$ as results. In turn, ``\textit{European football clubs in which a Brazilian player won a trophy}" expects triples (3-tuples) $<$\textit{club}, \textit{player}, \textit{trophy}$>$ as results. 

Each pair of entities $E_i$, $E_{i+1}$ in an entity tuple is connected with a relationship $R(E_i,E_{i+1})$. A complex information need can be expressed in a relational format, which is decomposed into a set of sub-queries that specify types of entities $E$ and types of relationships $R(E_i,E_{i+1})$ between entities. For each relationship query there is one query for each entity involved in the relationship. Thus a E-R query $Q$ that expects 2-tuples, is mapped into a triple of queries $(Q^{E_i}$, $Q^{R_{i,{i+1}}}$, $Q^{E_{i+1}})$, where $Q^{E_i}$ and $Q^{E_{i+1}}$ are the entity types for $E_i$ and $E_{i+1}$ respectively, and $Q^{R_{i,i+1}}$ is a relationship type describing $R(E_i,E_{i+1})$. For instance, ``football players who dated top models'' with answers such as $<$\textit{Cristiano Ronaldo}, \textit{Irina Shayk}$>$) is represented as three queries $Q^{E_i}=\{$\textit{football players}$\}$, $Q^{R_{i,i+1}}=\{$\textit{dated}$\}$, $Q^{E_{i+1}}=\{$\textit{top models}$\}$. 

Consequently, we can formalize that a query $Q$ contains a set of sub-queries $Q^{E} = \{Q^{E_1}, Q^{E_2}, ..., Q^{E_n} \}$ and a set of sub-queries $Q^{R} = \{Q^{R_{1,2}}, Q^{R_{2,3}}, ..., Q^{R_{n-2,n-1}} \}$. Automatic mapping of terms from a natural language information need $Q$ to queries $Q^{E_i}$ or $Q^{R_{i,i+1}}$ is out of the scope of this work and can be seen as a problem of query understanding \cite{pound2012interpreting}. We assume that the information needs are decomposed into constituent queries either by processing the original query $Q$ or by user input through an interface that enforces this structure $Q = \{Q^{E_i}$, $Q^{R_{i,i+1}}$, $Q^{E_{i+1}} \}$.

\section{Early Fusion}

E-R retrieval requires collecting evidence for both entities and relationships that can be spread across multiple documents. Therefore, it is not possible to create direct term-based representations. Documents serve as bridges between entities, relationships and queries. We propose an early fusion strategy specific to E-R retrieval that is inspired on the early fusion design pattern for object retrieval \cite{zhang2017design}. Therefore, our design pattern can be thought as creating a meta-document $D^{R_{i,i+1}}$ for each pair of entities (relationship) that co-occur close together in raw documents and a meta-document $D^{E_i}$ for each entity, similar to \textit{Model 1} of \cite{balog2006formal}. 

In our approach we focus on sentence level information about entities and relationships although the design pattern can be applied to more complex segmentations of text (e.g. dependency parsing). We rely on Entity Linking methods for disambiguating and assigning unique identifiers to entity mentions on raw documents $D$. We collect entity contexts across the raw document collection and index them in the \textit{entity index}. The same is done by collecting and indexing entity pair contexts in the \textit{relationship index}.  

We define the (pseudo) frequency of a term $t$ for an entity meta-document $D^{E_i}$ as follows:

\begin{equation} \label{tf_e}
f(t,D^{E_i}) =  \sum_{j=1}^{n} f(t,E_i,D_j) w(E_i,D_j)
\end{equation}

where $n$ is the total number of raw documents in the collection,  $f(t,E_i,D_j)$ is the term frequency in the context of the entity $E_i$ in a raw document $D_j$.  $w(E_i,D_j)$ is the entity-document association weight that corresponds to the weight of the document $D_j$ in the mentions of the entity $E_i$ across the raw document collection. Similarly, the term (pseudo) frequency of a term $t$ for a relationship meta-document $D^{R_{i,i+1}}$ is defined as follows:

\begin{equation} \label{tf_r}
f(t,D^{R_{i,i+1}}) =  \sum_{j=1}^{n} f(t,R_{i,i+1},D_j) w(R_{i,i+1},D_j)
\end{equation}

where $f(t,R_{i,i+1},D_j$ is the term frequency in the context of the pair of entity mentions corresponding to the relationship $R_{i,i+1}$ in a raw document $D_j$ and $w(R_{i,i+1},D_j)$ is the relationship-document association weight. In this work we use binary associations weights indicating the presence/absence of an entity mention in a raw document, as well as for a relationship. However, other weight methods can be used.

The relevance score for an entity tuple $T_E$ can then be calculated by summing the score of individual entity meta-documents and relationship meta-documents using standard retrieval models. Formally, the relevance score of an entity tuple $T_E$ given a query $Q$ is calculated by summing individual relationship and entity relevance scores for each $Q^{R_{i,i+1}}$ and $Q^{E_i}$ in $Q$. We use the terminology $|Q|$ to denote the number of sub-queries in a query $Q$. We define the score for a tuple $T_E$ given a query $Q$ as follows:

\begin{equation} \label{ef_eq}
  score(T_{E}, Q) =\sum_{i=1}^{|Q|-1} score(D^{R_{i,i+1}}, Q^{R_{i,i+1}})+\sum_{i=1}^{|Q|} score(D^{E_i}, Q^{E_i}) w(E_i, R_{i,i+1})
\end{equation}

where $score(D^{R_{i,i+1}}, Q^{R_{i,i+1}})$ represents the retrieval score resulting of the match of the query terms of a relationship (sub-)query $Q^{R_{i,i+1}}$ and a relationship (entity-pair) meta-document $D^{R_{i,i+1}}$. The same applies to the retrieval score $score(D^{E_i}, Q^{E_i})$ which corresponds to the result of the match of an entity (sub-)query $Q^{E_i}$ with a entity meta-document $D^{E_i}$. 

We use a binary association weight for $w(E_i, R_{i,i+1})$ which represents the presence of a relevant entity $E_i$ to a sub-query $Q^{E_i}$ in a relationship $R_{i,i+1}$ relevant to a sub-query $Q^{R_{i,i+1}}$. This entity-relationship association weight is the building block that guarantees that two entities relevant to (sub-)queries $Q^E$ that are also part of a relationship relevant to a (sub-)query $Q^R$ will be ranked higher than tuples where just one or none of the entities are relevant to the entity (sub-)queries $Q^E$.

For computing both $score(D^{R_{i,i+1}}, Q^{R_{i,i+1}})$ and $score(D^{E_i}, Q^{E_i})$ any retrieval model can be used. In this work we run experiments using Dirichlet smoothing Language Models (LM) and BM25. Considering Dirichlet smoothing unigram Language Models (LM) the scores can be computed as follows:
\begin{equation}
score_{LM}(D^{R_{i,i+1}}, Q^{R_{i,i+1}}) = \sum_{t}^{Q^{R_{i,i+1}}} \text{log}  \left (\frac{ f(t,D^{R_{i,i+1}}) +  \frac{f(t,C^R)}{|C^R|}\mu^R}{|D^{R_{i,i+1}}| + \mu^R} \right ) 
\end{equation}

\begin{equation}
score_{LM}(D^{E_i}, Q^{E_i}) = \sum_{t}^{Q^{E_{i}}} \text{log} \left (\frac{f(t,D^{E_i})
 + \frac{ f(t,C^E)}{|C^E|}\mu^E }{|D^{E_i}| + \mu^E} \right )
\end{equation}

where $t$ is a term of a (sub-)query $Q^{E_i}$ or $Q^{R_{i,i+1}}$, $f(t,D^{E_i})$ and
$f(t,D^{R_{i,i+1}})$ are the (pseudo) frequencies defined in equations \ref{tf_e} and \ref{tf_r}. The collection frequencies $f(t,C^E)$, $f(t,C^R)$ represent the frequency of the term $t$ in either the \textit{entity index} $C^E$ or in the \textit{relationship index} $C^R$. $|D^{E_i}|$ and$|D^{R_{i,i+1}}|$ represent the total number of terms in a meta-document while $|C^R|$ and $|C^E|$ represent the total number of terms in a collection of meta-documents. Finally, $\mu^E$ and $\mu^R$ are the Dirichlet prior for smoothing which generally corresponds to the average document length in a collection. 

Using BM25, the score is computed as summation over query terms, as follows:

\begin{equation}
score_{BM25}(D^{R_{i,i+1}}, Q^{R_{i,i+1}}) = \sum_{t}^{Q^{R_{i,i+1}}}\frac{ f(t,D^{R_{i,i+1}})(K_1 + 1)} {  f(t,D^{R_{i,i+1}}) + K_1 (1 - b + b \frac{|D^{R_{i,i+1}}|}{avg(D^{R})})} IDF(t)
\end{equation}

\begin{equation}
score_{BM25}(D^{E_i}, Q^{E_i}) = \sum_{t}^{Q^{E_{i}}}\frac{ f(t,D^{E_i}) (K_1 + 1)} {  f(t,D^{E_i}) + K_1 (1 - b + b \frac{|D^{E_i}|}{avg(|D^{E}|)})} IDF(t)
\end{equation}

where $IDF(t)$ is computed as log$\frac{N - n(t) + 0.5}{n(t) + 0.5}$ with $N$ as the number of meta-documents on the respective collection and $n(t)$ the number of meta-documents where the term occurs. $|D^{E_i}|$ and $|D^{R_{i,i+1}}|$ are the total number of terms in a meta-document, while $avg(|D^{E}|)$ and $avg(D^{R})$ are the average meta-document lengths. $K_1$ and $b$ are free parameters usually chosen as 1.2 and 0.75, in the absence of specific optimization.

\section{Experimental Setup}

\subsection{Test Collections}

We ran experiments with a total of 469 E-R queries aiming for 2-tuples of entities as results. We leave experimentation with longer E-R queries (e.g. 3-tuples) for future work. Relevance judgments consist of pairs of entities linked to Wikipedia.

Query sets for E-R retrieval are scarse. Generally entity retrieval query sets are not relationship-centric \cite{yahya2016relationship}. To the best of our knowledge there are only 3 test collections specifically created for E-R retrieval: ERQ \cite{li2012entity}, COMPLEX \cite{yahya2016relationship} and RELink \cite{saleiro2017relink}.  Neither ERQ nor COMPLEX provide complete relevance judgments and consequently, we manually evaluated each answer in our experiments.

ERQ consists of 28 queries that were adapted from INEX17 and OWN28 initiatives. Twenty two of the queries express relationships, but already have one entity instance named and fixed in the query (e.g. \textit{``Find Eagles songs''}). Only 6 queries ask for pairs of unknown entities, such as ``\textit{Find films starring Robert De Niro and please tell directors of these films.}''. 

COMPLEX queries were created with a semi-automatic approach. For a specific domain in a knowledge graph, a pivot entity is selected based on prior domain popularity. A chain of 2-4 entities connected to the entity is created based on a number of facts connecting to the pivot table. A set of different chains from several domains was given to human editors to formulate E-R queries answered by the entities in each chain. The query set contains 70 queries from which we removed 10 that expect 3-tuples of entities. COMPLEX consists of pure relationship-centric queries for unknown pairs of entities, such as ``\textit{Currency of the country whose president is James Mancham} ``\textit{Kings of the city which led the Peloponnesian League.}''  and ``\textit{Who starred in a movie directed by Hal Ashby?}''.

RELink queries and relevance judgments were also created with a semi-automatic approach. A sample of relational tables from Wikipedia was used as input to human editors for manually creating E-R queries. Columns from selected tables represent entity types and the table structure implies one or more relationships among the entities. Relevance judgments are automatically collected from each table. RELink comprises 600 queries aiming 2-tuples and 3-tuples of entities from which we use the subset of 381 queries aiming for pairs of related entities as results.

\subsection{Data and Indexing}

We aim to answer E-R queries without specific or pre-defined entity or relationship types. Therefore we use unstructured texts mentioning entities and relationships between entities to create our indices.
We use a dump of English Wikipedia from October 2016 and the ClueWeb-09-B\footnote{\url{http://www.lemurproject.org/clueweb09/}} collection combined with FACC1\cite{gabrilovich2013facc1} text span annotations with links to Wikipedia entities (via Freebase). The entity linking precision and recall in FACC1 are estimated at 85\% and 70-85\%, respectively \cite{gabrilovich2013facc1}. 

For our experiments we create two main indices: one for entity extractions and one for entity pairs (relationships) extractions. For a given Wikipedia article representing an entity we index each sentence and consider it as an entity occurrence extraction in the entity index. The Wikipedia dump used contains 4.1M entities. We use ClueWeb-09-B corpus with FACC1 annotations to extract relationship occurrences using an Open Information Extraction method like \cite{schmitz2012open}. We look for co-occurring entities in the same sentence of ClueWeb-09-B and we extract the separating string, i.e., the context of the relationship connecting them. We obtained 418M entity pairs extractions representing 71M unique entity-relationships. We ran our experiments using Lucene and made use of GroupingSearch for grouping extractions by entity and entity pair on query time.

\subsection{Retrieval Method}

We adopted a two stage retrieval approach. First, queries $Q^{E_i}$,$Q^{E_{i+1}}$ are submitted against the entity index and $Q^{R_{i,i+1}}$ is submitted against the entity-pair index. Initial sets of top 20K results grouped by entity or entity-pairs, respectively, are retrieved using Lucene's default search settings. Second, the score functions of the specific retrieval model are calculated for each set, using an in-house implementation. This process is easily parallelized. The final ranking score for each entity-pair is then computed using the early fusion strategy equation for $score(T_{E}, Q)$. 

We do not optimize the Dirichlet priors $\mu^E$ and $\mu^R$ in language models and set them equal to the average entity and relationships extractions length, respectively. The same happens with $K_1$ and $b$ in BM25, set to default values of 1.2 and 0.75, respectively. Evaluation scores are reported on the top 100 entity-pair results.

\section{Results}

\begin{table}[h]
\centering
\caption{Results of early fusion strategy using LM and BM25 on different query collections.}
\label{res}
\begin{tabular}{|l|l|l|l|l|}
\hline
          & \multicolumn{4}{c|}{\textbf{ERQ}}              \\ \hline
          & MAP    & P@10     & NDCG@10     & MRR \\ 
LM   & 0.1345  & 0.081   & 0.1468 & 0.1810  \\ 
BM25      & 0.1254 & 0.089   & 0.1563 & 0.1596  \\ \hline
          & \multicolumn{4}{c|}{\textbf{COMPLEX}}          \\ \hline
          & MAP    & P@10     & NDCG@10     & MRR \\ 
LM   & 0.1455  & 0.0567   & 0.1702 & 0.1437  \\ 
BM25      & 0.1223 & 0.049   & 0.1497 & 0.1416  \\ \hline
          & \multicolumn{4}{c|}{\textbf{RELink}}           \\ \hline
          & MAP    & P@10     & NDCG@10     & MRR \\ 
LM   & 0.0221  & 0.0084   & 0.0254 & 0.0260  \\ 
BM25      & 0.0229 & 0.0078   & 0.0247 & 0.0255  \\ \hline
\end{tabular}
\end{table}

We present the results of our experiments in Table \ref{res}. We report scores of four different retrieval metrics: Mean Average Precision at 100 results (MAP), precision at 10 (P@10),  normalized discounted cumulative gain at 10 (NDCG@10) and mean reciprocal rank (MRR). 
The first observation is concerning the retrieval model (LM vs BM25). On ERQ, LM shows higher MAP and MRR while BM25 has higher scores for metrics at top 10 results (P@10 and NDCG@10). Although results of both retrieval models are similar, LM outperforms BM25 for every metric on COMPLEX query collection. BM25 has higher MAP on RELink but it is lower on the remaining metrics. 

The second observation is concerned with the RELink results which are far lower for both retrieval models on all metrics. The RELink collection is by far the largest collection from the 3, comprising a total of 381 queries. It contains several queries regarding dates. For instance, the query ``Find australian films of 1981 and their directors.'' returns several entity-pairs comprising australian films and directors of those films but not from 1981. The most common relationship query $Q^R$ in this collection is ``located in'' which is a very frequent relationship string in our entity-pair index. We hypothesize that returning 20k entity-pairs on the first passage might result insufficient for RELink as it reduces the search space. In the future, we will further experiment with higher number of results.

\section{Concluding Remarks}

Work reported in this paper is concerned with expanding the scope of entity-relationship search methods to enable search over large corpora with flexible entity types and complex relationships. We have presented an early fusion strategy for fusion-based E-R retrieval. We anticipate that such strategy can be used as flexible baseline for further experimentation.  
For the sake of simplicity and clarity, we have reported on the basic E-R retrieval comprising a single relationship between two entities. In future work, we will report experiments with multiple relationships, as well as, an alternative late fusion strategy for E-R retrieval.

\bibliographystyle{unsrt}
\bibliography{refs} 

\end{document}